# Usage-based vs. Citation-based Methods for Recommending Scholarly Research Articles

André Vellino
School of Information Studies
University of Ottawa
avellino@uottawa.ca

## ABSTRACT

There are two principal data sources for collaborative filtering recommenders in scholarly digital libraries: usage data obtained from harvesting a large, distributed collection of Open URL web logs and citation data obtained from the journal articles. This study explores the characteristics of recommendations generated by implementations of these two methods: the 'bX' system by ExLibris and an experimental citation-based recommender, Sarkanto. Recommendations from each system were compared according to their semantic similarity to the seed article that was used to generate them. Since the full text of the articles was not available for all the recommendations in both systems, the semantic similarity between the seed article and the recommended articles was deemed to be the semantic distance between the journals in which the articles were published. The semantic distance between journals was computed from the "semantic vectors" distance between all the terms in the full-text of the available articles in that journal and this study shows that citation-based recommendations are more semantically diverse than usage-based ones. These recommenders are complementary since most of the time, when one recommender produces recommendations the other does not.

## Categories and Subject Descriptors

H.3.3 [**Information Systems**]: Information Search and Retrieval

## 1. INTRODUCTION

Recommender systems that apply collaborative filtering algorithms on user-ratings have been in existence for almost twenty years and are now used commercially to recommend a wide variety of products such as music, books and movies as well as people and metadata in social networking sites.

In digital libraries and publishers' portals, recommender systems are being adopted in the hope that they will help users to discover relevant "related literature" that may not have any keywords in common with content for which the user was searching. Several digital library recommenders have been studied or deployed including TechLens+ [15], BibTip [11] and CiteULike [5]. Each of these systems uses different kind of data from which to generate recommendations: TechLens+ uses both citation data and usage data, BibTip uses only OPAC usage data, and CiteULike harvests users' bookmarked items.

These collaborative filtering recommenders operate either by clustering similar items according to some characteristic of the item (e.g. text-content) or by profiling users' behaviour and clustering users according to some measure of similarity among them (user-based collaborative filtering) or by combining the two (hybrid recommenders.) The usage data with which to perform collaborative filtering is either co-download or co-citation information whereas content-based recommenders apply some measure of similarity between one item with all the other items using the salient content features can be either extracted from the items or obtained from the metadata such as bibliographic categories, authors, title, abstract, etc. It is known from these and other experiments that item-based collaborative filtering does not produce useful recommendations if the usage data is sparse. This occurs in situations where the ratio of end-users to the number of articles is low and the ratio of retrieved articles to the total number of articles is also low.

A disadvantage of content-based methods is that their recommendations will rarely stray far from the content-clusters of the previously rated items. Methods to compensate for the over-specialization of content-based recommendations include introducing randomness in the recommendation and filtering out the items that are *too* similar, and complementing them with collaborative filtering, which provide a source of naturally occurring serendipity from user behaviour [19]. This study, first reported on briefly in [16], compares two types of recommenders: the 'bX' recommender by Ex Libris that applies collaborative filtering to distributed usage data and a variant of Synthese [17] called Sarkanto that uses only citation data. The objective of this study was to compare and contrast some of the recommendation result-sets and behavioural characteristics of these two approaches: how each approach deals with the sparsity problem §2 and how much semantic diversity there is among the recommendations produced §3. Section §4 interprets the results and section §5 discusses future avenues of research.

## 2. DATA SPARSITY

One of the central problems for all recommender systems, but especially for those in digital libraries, is the sparsity of data they use to generate recommendations. If we measure the sparsity of a user-item matrix as the number of links between users and items (either ratings, or the occurrence of a download or citation) divided by the total number of possible links between users and items, then the sparsity of data used for typical collaborative filtering tasks, such as the Netflix data set [4], is $1.17 * 10^{-2}$. However, in a digital library these ratios are orders of magnitude smaller than in recommenders for commercial merchandise – on the order of tens of thousands of users per month for a collection of tens of millions of items. For example, the sparsity of the matrix of (scholars) to items (articles) in a substantial bibliographic portal such as Mendeley [9] is $2.66*10^{-5}$, almost three orders of magnitude smaller than Netflix. These extremely small user-item ratios are insufficient for collaborative filtering to produce reliable recommendations.

'bX' and Sarkanto are instances of two different strategies for overcoming the data sparsity problem. These strategies also yield very different kinds of recommendations that are complementary in several respects. In the next sections we describe the strategies and compare their merits.

### 2.1 'bX'

The 'bX' recommender is a web-service, commercially available from ExLibris [1], that recommends journal articles using data obtained from OpenURL logs of users' co-downloads. The recommender's design is based on research on the large scale usage of scholarly resources by Bollen and van de Sompel [6] that permits the harvesting of inter-institutional aggregation of log data. The quantity of data obtained with this method is sufficiently large to make it possible to apply a collaborative filtering algorithm [3] that predicts recommended articles in much the same manner as Netflix recommends movies or iTunes recommends songs. As OpenURL resolver logs grow over time, 'bX' recommendations reflect users' aggregate behavior with increasing precision and accuracy [10].

### 2.2 Sarkanto

Another strategy that has been used to address data sparsity, e.g. by TechLens+ [15], is to take advantage of bibliographic citations in the articles as a substitute for user ratings. The basic idea is to consider an article as a "user" and view the articles that it cites to be the article's "preferences" (or boolean ratings). The prototype Sarkanto recommender [2] uses a similar citation-based strategy to TechLens+ for addressing the sparsity problem and also enables the user to compare its recommendations side by side with the results from 'bX'. The Synthese architecture from which Sarkanto is derived can also incorporate other kinds of data, such as how users co-download (as in 'bX') but also how similar documents are to one another (content-based filtering) and explicit ratings submitted by users. For this experiment, Synthese, which was implemented with the Apache Mahout recommender, was used in a citation-data only mode.

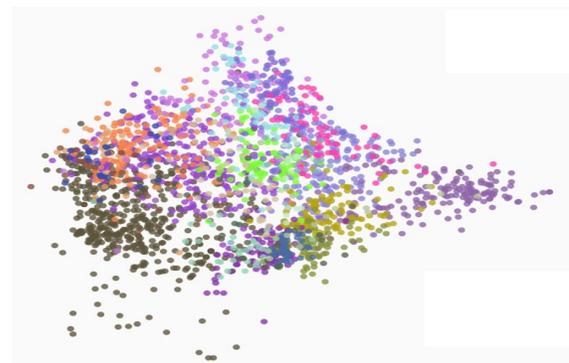

**Figure 1: Journal Semantic Distance Map in Newton, Callahan and Dumontier [12]. Different colours indicate journals in different fields**

## 3. COMPARISON

A typical method for assessing the quality of a recommender gauges the effectiveness of an algorithm by comparing known-to-be-good recommendations with algorithmically predicted ones [10]. For a recommender of scholarly articles that uses citation-based "ratings" such as Sarkanto, a sample of test articles is selected and, for each article in that set, one reference is removed and the recommender is tested for whether it predicts the removed reference. If the removed reference ranks highest in the list of recommendations, it belongs to the Top-1 recommendations, if it ranks in the first five recommendations it belongs to the Top-5, etc.

By this measure the accuracy rate of Sarkanto is in the vicinity of 20% [14], which is a respectable score given the limitations of the citation data in this study (only 25% of the articles in this collection have citation data available). However, the measures often used to evaluate algorithms such as "Top-N" (above) or Mean Absolute Error (MAE) [10] are not applicable in this situation because the Open URL log data do not enable the tester to determine which are the "known-to-be-good" recommendations. Hence, a meaningful comparative evaluation of the quality of recommendations generated by each recommender could only be provided by a human-subject expert that inspects the results and assesses the relevance of each recommendation.

Therefore, instead of using any of the above measures (MAE, Top-N, or human subject studies), we compared how each of these two strategies behave with respect to quantity, diversity and complementarity.

### 3.1 A Priori Comparison

The Sarkanto and 'bX' recommenders each have *a priori* strengths and weaknesses in their respective approaches. For instance, while 'bX' can take advantage of a voluminous amount of usage data distributed across the world, this data may not reflect, even in the aggregate, the interests of specialists in any given field. Usage data from OpenURL logs is indiscriminate between expert researchers and undergraduate university students. In addition, a dependence on usage information makes such a recommender unable to address the recommendation needs of users interested in the end of "long tail" of sparsely researched areas.

On the other hand a recommender that uses bibliographic citations instead of usage data suffers from other limitations. One is that while citations are an indicator of relevance, they are not necessarily an indication of *favourable* relevance in the mind of the author. Findings in [7] showed that authors were motivated to cite a work for a variety of reasons, including the fact that it might promote the authority of their own work and that the cited work deserved criticism. Another limitation is that citations are static over time and citation-based recommendations won't reflect current usage trends.

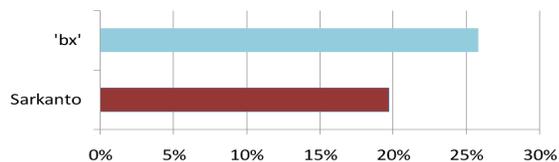

Figure 2: Recommendation coverage

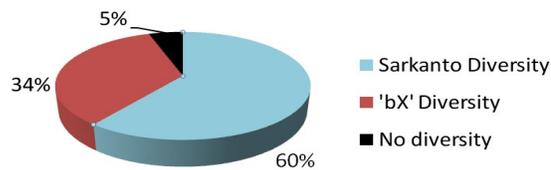

Figure 3: Diversity of recommendations

### 3.2 Experimental Comparison

In this experiment we compared the semantic diversity of recommendations, the number of recommended articles and the extent to which these two approaches produced any recommendations at all. End users often want recommendations to be some "semantic distance" away from what they are searching for yet still be relevant. One major function of a recommender is to propose related items that do not necessarily share terms in common with the items that the user has already found.

The semantic distance between the seed article and the recommendations generated by both 'bX' and Sarkanto was measured by examining the "journal diversity" among the recommended articles relative to the original article. It would have been preferable to use the semantic distance between the seed article and each of the recommended articles based on their full-text content. However, in the absence of the full-text for all of the recommended articles, we used an aggregate measure: the semantic map produced by Newton et. al. [12] that calculates the semantic distances between 2365 journals, reproduced in Figure 1. On this map, a 'journal' is considered to be the concatenation of the full-text of all the available articles in that journal.

The journal-to-journal distances in this map were computed using Widdows' Semantic Vectors method [18] applied to the full text of a collection of 5.7 million articles. The average distance between a randomly selected pair of such journals is 0.7875 where 1.0 is the distance between a journal and itself and 0.0 is the distance between two journals that have no terms in common. In this collection, the two most similar journals are at a distance of 0.998 and the two most dissimilar are at a distance of 0.2724.

From a subset of approximately 2 million test articles that contain about 11.8 million references, we randomly selected a statistically significant sample of 1886 articles that were used as seeds for generating recommendations. For each seed article, we compared the recommendations generated by 'bX' and Sarkanto and counted:

- the number of recommended articles
- the semantic distance between the seed article and the recommended articles
- the number of times that 'bX' and Sarkanto both recommended articles from a given seed article
- for each instance where both 'bX' and Sarkanto produced a set of recommendations from the same seed, which one of 'bX' or Sarkanto had greater journal diversity

Note that the variety of journals that can be recommended in the 'bX' system is significantly greater than the range available in Sarkanto, given the limited number of publishers (about 50) in the article collection used by Sarkanto.

### 4. RESULTS

Out of the 1886 test runs, 487 generated one or more recommendations using 'bX' compared with 372 for Sarkanto (i.e. the recommendation coverage was 26% vs. 20% in Figure 2). However, on average, Sarkanto recommended more articles per seed-article and the total number of recommended articles for each system was about the same (51% vs 49% in Figure 4). The relatively low coverage of citation-based recommendations is an artefact of the distribution of citation information available in the collection.

The number seed articles that produced recommendations with both Sarkanto and 'bX' was only 13% (116 out of a total of 859). This means that most of the time either one or the other recommender would produce a result, indicating a high degree of complementarity. Furthermore, within this 13% of articles for which both recommenders produced a result, none of the recommended articles were the same. Using these recommenders together increased the coverage of recommendations to 60%.

Figure 3 shows the relative number of recommendations that were more semantically diverse in Sarkanto than in 'bX' from among the set of seed articles that produced joint results. About 5% of the time, both recommenders produced results with no semantic diversity (i.e. all the recommended articles came from the same journal as the seed article).

A qualitative inspection of this set of joint results shows that 'bX' recommendations tend to be more recent than Sarkanto's, which is easily explained by the lag time needed for published articles to receive any citations (2 years) compared with co-download data [13]).

Overall, the average semantic distance between the seed article and the recommendations was 0.906 for Sarkanto and 0.935 for 'bX'.

### 5. DISCUSSION AND FUTURE WORK

The lesser diversity of 'bX' recommendations can be explained by the likelihood that co-downloaded articles were obtained from a list of search results. A manual inspection of the 13% of jointly occurring recommendations results showed that both 'bX' and Sarkanto generated recommendations that were topical. For instance, if the source article was about angioplasty the recommended articles tended to also be about either atherosclerosis or coronary artery disease. Recommendations from 'bX', however, tended to contain similar keywords to words occurring in the title of the seed article. The search-engine genesis hypothesis could be tested by examining OpenURL logs and the search results that led to the invocation of the Open URL resolver.

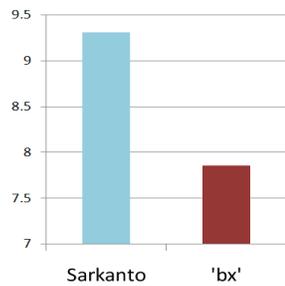

**Figure 4: Number of recommendations per article**

The low percentage of articles which yield recommendations with both recommenders reflects their complementarity. The user-based method recommends "other articles that other users also downloaded" whereas the citation-based one recommends "other articles whose citation patterns are similar to this one".

One likely consequence of employing a usage-only method of recommending articles is that, as researchers download recommended articles, the OpenURL logs generated by such usage will begin to self-reinforce these same recommendations in the future (see [8] for a discussion of this effect). The resulting recommendations would then converge to be the equivalent of "Top-N" downloads, which may considerably reduce the diversity of recommendations.

It would be valuable to undertake a human-subject study that compares the usefulness of a system that converges to this fixed point over time with one that has already achieved a fixed point from (relatively) static citation patterns. In addition, it is likely that the extent of semantic diversity among recommendations from both methods varies considerably across disciplines. Running these same experiments with articles seeded from journals only in given disciplines would measure the semantic diversity of recommendations in different disciplines.